\documentclass[letterpaper, 10 pt, conference]{ieeeconf}  

\IEEEoverridecommandlockouts                              
                                                          
\usepackage{amsmath,amssymb}
\usepackage{xfrac}
\usepackage{mathtools,cuted}

\usepackage{bm}

\usepackage{algpseudocode}
\usepackage{tabularx}
\usepackage[final]{pdfpages}
\usepackage[ruled,vlined,noend,linesnumbered]{algorithm2e}
\usepackage{booktabs}
\usepackage[short]{optidef}

\usepackage{tablefootnote}

\usepackage{physics}

\usepackage{tikz}
\usepackage{svg}
\usepackage{placeins}
\usepackage{units}
\usepackage{arydshln}
\usepackage{graphicx}
\usepackage{soul,xcolor}
\usepackage{caption}
\usepackage{subcaption}
\usepackage{float}

\usepackage{url}
\usepackage{hyperref}
\usepackage{cite}
\usepackage[capitalise]{cleveref}

\usepackage{comment} 

\newcommand{\argmin}{{\mathrm{argmin}}}
\newcommand{\argmax}{{\mathrm{argmax}}}

\newcommand*{\minOp}{\operatornamewithlimits{min}\limits}

\newcommand{\GP}{\mathcal{G\!P}}

\newcommand{\transp}{\mathsf{\scriptscriptstyle T}}

\iftrue
\newcommand{\eye}{\mathbb{I}}
\else
\newcommand{\eye}{\mathbf{I}}
\fi
\newcommand{\vc}[1]{{ \mathrm{#1} }}
\newcommand{\mx}[1]{{ \mathrm{#1} }}

\newcommand{\nth}{{\text{\tiny{th}}}}

\newcommand{\Rbb}{\mathbb{R}}
\newcommand{\Ncal}{\mathcal{N}}

\newcommand{\smtiny}[1]{{\scalebox{.65}{#1}}}

\newcommand{\pref}{p_{\text{ref}}}
\newcommand{\vref}{v_{\text{ref}}}

\newcommand{\rd}{\boldsymbol{r_d}} 
\newcommand{\init}{{\text{init}}} 
\newcommand{\Dcal}{{\mathcal{D}}} 
\newcommand{\bxi}{{\boldsymbol{\xi}}} 
\newcommand{\btheta}{{\boldsymbol{\theta}}} 

\newcommand{\sss}[1]{{\smtiny{(#1)}}} 

\begin{document}
\title{Performance-based Trajectory Optimization for Path Following Control Using Bayesian Optimization}
\author{Alisa Rupenyan$^{1,2*}$, Mohammad Khosravi$^{1,*}$, John Lygeros$^{1}$
\thanks{This project was funded by the Swiss Innovation Agency, grant Nr. 46716, and by the Swiss National Science Foundation under NCCR Automation.}
\thanks{$^{1}$ Automatic Control Laboratory, ETH Zurich, Switzerland }
\thanks{$^{2}$ Inspire AG, Zurich, Switzerland }
\thanks{$^*$ The authors contributed equally.}
}
\maketitle
\begin{abstract}

Accurate positioning and fast traversal times determine the productivity in machining applications. This paper demonstrates a hierarchical contour control implementation for the increase of productivity in positioning systems. The high-level controller pre-optimizes the input to a low level cascade controller, using a contouring predictive control approach. This control structure requires tuning of multiple parameters. We propose a sample-efficient joint tuning algorithm, where the performance metrics associated with the full geometry traversal are modelled as Gaussian processes and used to form the global cost and the constraints in a constrained Bayesian optimization algorithm. This approach enables the trade-off between fast traversal, high tracking accuracy, and suppression of vibrations in the system. The performance improvement is evaluated numerically when tuning different combinations of parameters. We demonstrate that jointly tuning the parameters of the contour- and the low level controller achieves the best performance in terms of time, tracking accuracy, and minimization of the vibrations in the system.
 
\end{abstract}
\section{Introduction}
\label{sec:introduction}


In machining, positioning systems need to be fast and precise to guarantee high productivity and quality. Such performance can be achieved by model predictive control (MPC) approach tailored for tracking a 2D contour \cite{MPCC2010, Liniger2019realtime}, however requiring precise tuning and good computational abilities of the associated hardware. Using the superior planning capabilities of MPC can be beneficial to generate offline an optimized trajectory which is provided as an input to a low level PID controller \cite{bx2}. While this approach improves the tracking performance of the system without imposing any requirements for fast online computation, it is associated with tuning of multiple parameters, both in the MPC planner, and in the low level controller. We propose an approach to automate the simultaneous tuning of multiple design variables using data-driven performance metrics in such hierarchical control implementations.

MPC accounts for the real behavior of the machine and the axis drive dynamics can be excited to compensate for the contour error to a big extent, even without including friction effects in the model \cite{bx1, Stephens2013}. High-precision trajectories or set points can be generated prior to the actual machining process following various optimization methods, including MPC, feed-forward PID control strategies, or iterative-learning control \cite{Tang2013,bx5}, where friction or vibration-induced disturbances can be corrected. In MPC, closed-loop performance is pushed to the limits only if the plant under control is accurately modeled, alternatively, the performance degrades due to imposed robustness constraints. Instead of adapting the controller for the worst case scenarios, the prediction model can be selected to provide the best closed-loop performance by tuning the parameters in the MPC optimization objective for maximum performance \cite{lu2020mpc,Piga_2019, sorourifar2020data}. Using Bayesian optimization-based tuning for enhanced performance has been further demonstrated for cascade controllers of linear axis drives, where data-driven performance metrics have been used to specifically increase the traversal time and the tracking accuracy while reducing vibrations in the systems \cite{khosravi2020performance,Khosravi2020Cascade}. The approach has been successfully applied to linear and rotational axis embedded in grinding machines and shown to standardize and automate tuning of multiple parameters \cite{konig2020safety}.

In this work, we use the model predictive contouring control (MPCC)
which is an MPC-based contouring approach to generate optimized tracking references. We account for model mismatch by automated tuning of both the MPC-related parameters and the low level cascade controller gains, to achieve precise contour tracking with micrometer tracking accuracy. The MPC-planner is based on a combination of the identified system model with the contouring terms. In our approach the tracking error is coupled with the progression along the path through the cost function. The automated tuning of the parameters is performed using a cost that accounts for the global performance over the whole trajectory. Additional constraints in the Bayesian optimization algorithm allow for balancing traversal time, accuracy, and minimization of oscillations, according to the specific crucial requirements of the application. We demonstrate enhanced performance in simulation for a 2-axis gantry, for geometries of different nature.

\section{Parameter Tuning in Contouring Control}
\label{sec:problem}

\begin{figure}[t]
    \centering
    \includegraphics[width=0.45\textwidth]{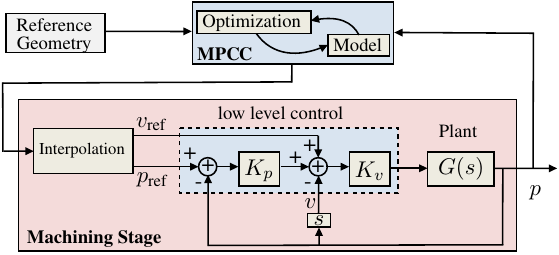}
    \caption{The model predictive contouring control (MPCC)}
    \label{fig:MPCC}
\end{figure}

In this paper, we focus on the biaxial machine tool contouring control problem, relevant for a production of multiple copies of a part with desired geometry $\rd:[0,L]\to\mathbb{R}^2$, parameterized by the arc-length, is expected to be traversed in minimum time while staying within a tolerance band perpendicular to the contour. The maximum allowed deviation given by the infinity norm of the tracking error should be smaller than $20$\,$\mu$m.
The machine has separate limitation for velocity, $v$, and acceleration, $u$, in both $x$ and $y$ directions. More precisely, we have $v \in \mathcal{V} := \{[v_x,v_y]^\transp| \, |v_x| \leq 0.2\text{m/s},\,|v_y| \leq 0.2\text{m/s}\}$ and $u\in \mathcal{U} := \{[u_x,u_y]^\transp|\,|u_x| \leq 20\text{m/s}^2,\,|u_y| \leq 20\text{m/s}^2\}$. 
The system is equipped with a machining stage unit where a cascade PI controller is employed (see Figure \ref{fig:MPCC}), and has a MPC-based planning unit (also noted as MPCC) to pre-optimize the references (position, velocity) provided for each geometry. 
The goal is to tune the parameters of the MPC-based planning unit without introducing any modification in the structure of the underlying control system.
We leverage the repeatability of the system, which is higher than the integrated encoder error of $3 \mu m$,
and introduce a parameter tuning strategy for the MPCC planner and of the low level controller gains using an automated data-driven approach, where the global performance metrics are computed from the full geometry traversal. 

\section{Plant and low level Control} \label{sec:plant_and_lowlevelcontrol}
The physical system is a 2-axis gantry stage for $(x,y)$ positioning with industrial grade actuators and sensors \cite{lanz2018efficient}. 
The plant can be modeled as a mass-spring-damper system with two masses linked with a damper and a spring for capturing imperfection and friction in the transmitting movement \cite{Qian2016NovelSM}. 
Let  $m_1$ and $m_2$ denote the mass of the moving parts with respective location $z = (x_1,y_1)$ and $p = (x_2,y_2)$, and movement friction coefficient $b_1$ and $b_2$. Also, let $k$ be the stiffness coefficient, $c$ be the damping coefficient, and $u = (u_1,u_2)$ be the force applied by the motor. 
From first principles, one has
\begin{equation}
\begin{split}
m_1\ddot{z} & = -b_1\dot{z} +  k(p - z) + c(\dot{p} - \dot{z}) + u(t),\\ 
m_2\ddot{p} & =  -b_2\dot{p} +  k(z - p) + c(\dot{z} - \dot{p}).
\end{split}
\label{eq:newton}
\end{equation}
Note that here $p= (x_2,y_2)$ corresponds to the contouring location; for ease of notation, we drop the index later and simply write $(x,y)$.
We model the system as two uncoupled axis with identical parameters. According to \eqref{eq:newton}, the plant can be described by the transfer function $G(s)$, from the force input to the the position of system, $p$, defined as 
\begin{equation}\label{eq:tf}
G(s)= \frac{cs+ k}{a_4s^4+ a_3s^3+ a_2s^2+ a_1s}\, ,   
\end{equation}
where $a_1= k(b_1+b_2)$, $a_2 = k(m_1+m_2) + c(b_1+b_2) + b_1b_2$, $a_3 = m_1(b_2 + c) + b_2(b_1 + c)$, and
$a_4 = m_1m_2$, $m_1$ and $m_2$ are known.
The rest of the parameters $w:=(b_1,b_2,c,k)$ have to be estimated. 

Let $p_{w}$ be the position trajectory obtained by simulating the corresponding dynamics with respect to the parameters $w$, and $J$ be the error metric between the real measurement of position $p$ and the simulated position $p_{w}$ defined as $J(w):=\|p-p_{w}\|_{\infty}+\|p-p_{w}\|_{2}$.
The error metric $J$ is introduced to consider the overshoots via the infinity norm, and also, the offset via the 2-norm. We estimate parameters $w$ as $\argmin_{w}J(w)$, which can be obtained using the available solvers for nonlinear optimization.

In the low level control of the plant, a cascade controller is employed for tracking the position and velocity reference trajectories 
(see Figure \ref{fig:MPCC}). 
More precisely, the force input applied to the system is  $u = K_v(K_p e_p + e_v)$, where $(K_p,K_v)$ are the nominal gains of the controller, and $e_p$ and $e_v$ are the error signals respectively in tracking the desired position and velocity trajectories.
One can easily obtain the transfer function from the reference trajectories to the actual position and velocity as 
\begin{equation}
    \begin{bmatrix}p\\ v\end{bmatrix}
    =
    \frac{1}{H(s)}
    \begin{bmatrix}
    H_1(s) & H_2(s)\\ 
    sH_1(s) & sH_2(s)
    \end{bmatrix} 
    \begin{bmatrix}\pref \\\vref
    \end{bmatrix},
\label{eq:Matsys}
\end{equation}
where 
$p:=(x,y)$, $p_{\text{ref}}:=(x_{\text{ref}},y_{\text{ref}})$, 
$v:=(\dot{x},\dot{y})$, $v_{\text{ref}}:=(\dot{x}_{\text{ref}},\dot{y}_{\text{ref}})$, 
$H_1,H_2$ and $H$ are defined as $H_1(s) = K_vK_pG(s)$, $H_2(s) = K_vK_pG(s)$, and $H(s)= K_vsG(s) + 1 + K_vK_pG(s)$.
Given \eqref{eq:Matsys}, one can obtain a discrete time model with sampling time $T=2.5\mathrm{ms}$ as 
\begin{equation}\!\!\!\!\!\!\!\!\!
\left\{\!\!\!\!
\begin{array}{l}
\zeta^{(x)}_{k+1} \!=\! A_{x}\zeta^{(x)}_{k} \!+\! B_{x} \!
\begin{bmatrix}x_{\text{ref},k}\\\dot{x}_{\text{ref},k} \end{bmatrix}\!\!,
\\ 
\begin{bmatrix} x_{k} \\ \dot{x}_{k} \end{bmatrix} 
\!= C_{x}\zeta^{(x)}_{k},
\end{array}\right.\!\!
\!\!\!
\left\{\!\!\!\!
\begin{array}{l}
\zeta^{(y)}_{k+1}\!=\! A_{y}\zeta^{(y)}_{k}\!+\! B_{y} \!
\begin{bmatrix}y_{\text{ref},k}\\\dot{y}_{\text{ref},k} \end{bmatrix}\!\!,
\\ 
\begin{bmatrix} y_{k} \\ \dot{y}_{k} \end{bmatrix}
\!= C_{y}\zeta^{(y)}_{k},
\end{array}\right.\!\!\!\!\!\!\!\!\!\!
\label{eq:st_space_model}
\end{equation}
where 
$[x_{k}, \dot{x}_{k}]^\transp$, $[y_{k}, \dot{y}_{k}]^\transp$,
$[x_{\text{ref},k}, \dot{x}_{\text{ref},k}]^\transp$ and $[y_{\text{ref},k}, \dot{y}_{\text{ref},k}]^\transp$ are respectively sampled version of
$[x, \dot{x}]^\transp$, $[y, \dot{y}]^\transp$,
$[x_{\text{ref}}, \dot{x}_{\text{ref}}]^\transp$ and $[y_{\text{ref}}, \dot{y}_{\text{ref}}]^\transp$, and, 
$\zeta^{(x)}_{k}$ and $\zeta^{(y)}_{k}$ respectively denote the state vectors in the corresponding discrete-time dynamics.  


\section{Model Predictive Contouring Control}

%
%
Model predictive contouring control (MPCC) is a control scheme based on minimisation of a cost function which trades the competing objectives of tracking accuracy and traversal time by adjusting the corresponding weights in the cost function. We now introduce the main ingredients for a MPCC formulation. 

\begin{figure}[t]
\centerline{\includegraphics[width=0.20\textwidth]{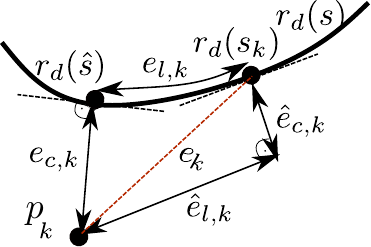}}
\caption{The error variables in the contouring MPC approach}
\label{fig:contour_errors}
\end{figure}

Let 
$p_{k} := (x_k,y_k)\in\mathbb{R}^2$ 
be the $k^\text{\tiny{th}}$ sample of the real position of system. 
Following \cite{Liniger2019realtime},
we approximate the \emph{true path parameter} 
$\hat{s}:= \argmin_{s\in[0,L]}\|\boldsymbol{r_d}(s)-p_{k}\|$ 
corresponding to 
$p_{k}$ 
by \emph{virtual path parameter}, $s_k$, defined as
\begin{align}
s_{k+1}=s_{k}+T v_{s,k} \, ,
\label{eq:path_param}
\end{align}
where $v_{s,k}$ is the sampled velocity along the path at time step $k$ and $T$ is the sampling time. 
The errors in this approximation, i.e., the deviations from the desired geometry, are characterized by the \emph{contour error},  $e_{c,k}$, and the \emph{lag error}, $e_{l,k}$ (see Figure \ref{fig:contour_errors}). Define the \emph{total error vector} as 
$\boldsymbol{e_k}=\boldsymbol{r_d}(s_k)-p_{k}$, 
and let  $\hat{e}_{l,k}$ and  $\hat{e}_{l,k}$ be the approximated errors defined respectively for $e_{l,k}$ and $e_{l,k}$ as shown in Figure \ref{fig:contour_errors}. 
Similar to \cite{Liniger2019realtime}, 
using linearization of the errors and Taylor expansion,  one can introduce the  following dynamics for the errors 
\begin{equation}\label{eq:3}
\begin{split}
&\hat{e}_{l,k+1}\!=\!\frac{{r'_{d,x}}(s_k)}{\| \boldsymbol{r'}(s_k) \|}(r_{d,x}(s_k)\!-\!x_k \!-\! T \dot{x}_k \!-\! \frac{1}{2}T^2 \ddot{x}_k) \\
&\ \ \  \!+\! \frac{{r'_{d,y}}(s_k)}{\| \boldsymbol{r'}(s_k) \|}(r_{d,y}(s_k)\!-\! y_k \!-\! T \dot{y}_k \!-\! \frac{1}{2}T^2\ddot{y}_k)\!+\! T v_{s,k}, 
\end{split}   \!\!\! 
\end{equation}
and
\begin{equation}\label{eq:controur_errors}
\begin{split}
\hat{e}_{c,k+1}\!=\!-\frac{{r'_{d,y}}(s_k)}{\| \boldsymbol{r'}(s_k) \|}(r_{d,x}(s_k)\!-\! x_k \!-\! T \dot{x}_k \!-\! \frac{1}{2}T^2 \ddot{x}_k)\\
+\frac{{r'_{d,x}}(s_k)}{\| \boldsymbol{r'}(s_k) \|}(r_{d,y}(s_k)\!-\!y_k \!-\! T \dot{y}_k \!-\! \frac{1}{2}T^2\ddot{y}_k)  \, , 
\end{split}    
\end{equation}
where $\boldsymbol{{r'}_d}(s):=(r'_{d,x}(s),r'_{d,y}(s))$ is the parametric derivative of $\boldsymbol{r_d}$, and 
$\ddot{x}_k,\ddot{y}_k$ are sampled acceleration of system in the $x$ and $y$ coordinates.  
When $\hat{e}_{l,k}$ is small, $\hat{e}_{c,k}$ approximates the contour error well, and the virtual path parameter is a good approximation of $\hat{s}$. 

To bring the model close to the real system, we unify the terms required for the contour control formulation with the velocity and acceleration for each axis from the identified, discretized state-space model from \eqref{eq:st_space_model}. 
Also, we include the path progress $s_k$ and the two error terms $\hat{e}_{l,k}$ and $\hat{e}_{c,k}$. Here, the velocities and accelerations correspond to the identified system dynamics \eqref{eq:st_space_model}. 
Accordingly, we introduce state vector $\mathbf{z}_k$ as
\begin{equation*}
\mathbf{z}_k:=[\zeta^{(x)}_k{}^\transp\  r_{x,k}\ \dot{r}_{x,k}\  \zeta^{(y)}_k{}^\transp\   r_{y,k}\  \dot{r}_{y,k}\ s_k\   \hat{e}_{l,k}\  \hat{e}_{c,k}]^\transp,    
\end{equation*}
where $r_{x,k}$, $\dot{r}_{x,k}$, $r_{y,k}$, and $\dot{r}_{y,k}$ are sampled version of $r_{d,x},\dot{r}_{d,x},r_{d,y}$, and $\dot{r}_{d,y}$, respectively.
Similarly, the inputs are collected in vector $\mathbf{u}_k$ defined as $\mathbf{u}_k := [ \ddot{x}_{k}\ \ddot{y}_k\ v_{s,k}]^\transp$. 
Following \eqref{eq:st_space_model}, \eqref{eq:path_param}, \eqref{eq:3} and \eqref{eq:controur_errors}, we obtain a linear time varying system of the form $\mathbf{z}_{k+1} = \mathrm{A}_k \mathbf{z}_k + \mathrm{B}_k \mathbf{u}_k + \mathbf{d}_k$, where only the error dynamics are time-dependent. 

The cost function of MPCC is designed to match the goals of the contouring controller for the biaxial system. We penalize the squared longitudinal error $\hat{e}_l^2$ since this error has to be small for the formulation to be accurate and the squared contouring error $\hat{e}_{c}^2$, as it corresponds to the tracking error. We reward progress at the end of the horizon $s_N$, which corresponds to traversing the geometry as fast as possible, and penalize the applied inputs (accelerations) and their change (jerk), to achieve smooth trajectories and minimize vibrations.
Accordingly, we can now formulate the trajectory optimization problem as
\begin{align}
\min_{\mathbf{Z},\mathbf{U}}  &\sum_{k=1}^{N-1}\!\!
\Big(\gamma_{l} \hat{e}_{l,k}^{\:2} + \gamma_{c} \hat{e}_{c,k}^{\:2} - \gamma_v v_{s,i} + \mathbf{u}_k^\transp R \mathbf{u}_k + \Delta_k \mathbf{u}^\transp S \Delta_k \mathbf{u}\Big) \nonumber\\&
\ \ +\gamma_{l,N} \hat{e}_{l,N}^{\:2} +\gamma_{c,N}, 
\hat{e}_{c,N}^{\:2}   -  \gamma_{s,N} s_N, 
\label{eqn:MPCC} 
\\
\mathrm{s.t.}\  
&\mathbf{z}_{k+1} = \mathrm{A}_k \mathbf{z}_k + \mathrm{B}_k \mathbf{u}_k + \mathbf{d}_k, 
\qquad\quad\ k=0,..,N-1,
\nonumber\\
&\hat{e}_{c,k} \in \mathcal{T}^c, \quad v_{k} \in \mathcal{V}, \quad u_k \in \mathcal{U}, 
\qquad k=0,..,N-1,
\nonumber\\
&v_{N} \in \mathcal{V}_N,\quad \hat{e}_{c,N} \in \mathcal{T}^c_{N}, \nonumber\label{eq:globalMPC}
\end{align}
where $\mathbf{z}_0$ is given, $\Delta_k \mathbf{u} := \mathbf{u}_{k} - \mathbf{u}_{k - 1}$ is the change in the acceleration,  $\mathbf{Z} := (\mathbf{z}_1,...,\mathbf{z}_N)$ and $\mathbf{U} := (\mathbf{u}_0,...,\mathbf{u}_{N-1})$ are the state and input trajectories, $\gamma_{l}$ and $\gamma_{c}$ are the error weights,  $R$ is a diagonal positive definite input weight matrix with non-zero terms $\gamma_{\tiny{\ddot{x}}}$ and $\gamma_{\tiny{\ddot{y}}}$ for the acceleration along the two axis, $S$ is a positive definite weight matrix applied to the acceleration differences with non-zero diagonal terms $\gamma_{\tiny{\dddot{x}}}$ and  $\gamma_{\tiny{\dddot{y}}}$, $\gamma_{v}$ is a linear reward for path progress and, $\mathcal{T}^c$ is the tolerance band of $\pm 20$\,$\mu$m,  $\gamma_{l,N}$ and $\gamma_{c,N}$ are terminal contouring weights, and $\gamma_{s,N}$ is the weight considered for the progress maximization.

\section{Data-driven Tuning Approach}
\label{sec:bo}
The overall performance of the introduced control strategy 
depends on the design parameters characterizing the MPCC scheme \eqref{eqn:MPCC} and the lower level control introduced in Section \ref{sec:plant_and_lowlevelcontrol}, i.e., the vector $\theta$ defined as
\begin{equation}
\label{eq:opt_var}
\theta :=
[\gamma_{c}, \gamma_{l}, \gamma_{\tiny{\ddot{x}}}, \gamma_{\tiny{\ddot{y}}}, \gamma_{\tiny{\dddot{x}}}, \gamma_{\tiny{\dddot{y}}}, \gamma_{v}, N,K_p,K_v]^\transp.
\end{equation}
These parameters implicitly encode trade-offs improving the accuracy of positioning, reducing the tracking time and reducing the impact of modeling imperfection. 
To explore these trade-offs we formulate high-level optimization problem with cost function and constraints defined based on the entire position and velocity trajectory, which indicate respectively the overall performance of the control scheme and the operation limits.

We have two main goals in the positioning problem: to traverse given geometry as fast as possible, and to adhere to the pre-defined tracking tolerances. The first goal is directly reflected by the total number of time steps necessary to traverse the whole geometry, denoted by $k_{\mathrm{tot}}$.
Note that $k_{\mathrm{tot}}$ depends implicitly on the vector of parameters $\theta$; accordingly, we define the cost function as $g_0(\theta):=k_{\mathrm{tot}}$.
For the second goal, aiming to minimize deviations and oscillations in the system, we introduce two constraints as
\begin{equation}
\begin{split}
&
g_1(\theta):=\max\Big(\big\|[\dddot{x}_k]_{k=1}^{k_{\mathrm{tot}}}\big\|_{\infty}, \big\|[\dddot{y}_k]_{k=1}^{k_{\mathrm{tot}}}\big\|_{\infty}\Big) 
\le q_{1,{\mathrm{tr}}},
\\&
g_2(\theta):=\big\|[\hat{e}_{c,k}]_{k=1}^{k_{\mathrm{tot}}}\big\|_{\infty} \le q_{2,{\mathrm{tr}}},
\end{split}
\end{equation}
where $\dddot{x}_k,\dddot{y}_k$  are sampled jerk of the system in the $x$ and $y$ coordinates, and $q_{1,{\mathrm{tr}}}$ and $q_{2,{\mathrm{tr}}}$ are threshold values for the maximal allowed jerk and the maximal allowed deviation along the whole geometry.
Note that the implicit dependency of these constraints to the vector of parameters is highlighted by the argument $\theta$ employed for $g_1$ and $g_2$.
The first constraint, $g_1(\theta) \le q_{1,{\mathrm{tr}}}$, limits oscillations in the system due to unaccounted changes in acceleration originating in the MPCC stage. The second constraint, $g_2(\theta)\le q_{2,{\mathrm{tr}}}$, limits the maximal allowed lateral error along the trajectory, and serves as an additional tuning knob on the tracking error. 
To tune the parameters $\theta$, we solve the following optimization problem 
\begin{equation}\label{eq:BO_min_f_st_g1g2}
\begin{array}{cl}
\minOp_{\theta\in\Theta} & g_0(\theta)\\
\text{s.t.} & g_i(\theta)\le q_{i,{\mathrm{tr}}},\quad i=1,2,
\end{array}
\end{equation}
where $\Theta$ is the feasible set for $\theta$. 
Note that, for a given $\theta\in\Theta$, the value of $g_0(\theta)$, $g_1(\theta)$ and $g_2(\theta)$ can be obtained by performing an experiment, where the design parameters are set to $\theta$, a complete cycle of operation of the system is performed, the values of the three functions are calculated based on the resulting full trajectory of position and velocity.
To reduce the number of times this experimental “oracle” is invoked, we employ Bayesian optimization (BO) \cite{shahriari2015taking,pmlr-v32-gardner14}, which is an effective method for controller tuning \cite{konig2020safety,khosravi2019controller,khosravi2019machine} and optimization of industrial processes \cite{Maier2019}. The constrained Bayesian optimization samples and learns both the objective function and the constraints online and finds the global optimum iteratively. For this, it uses Gaussian process regression \cite{Rasmussen} to build a surrogate for the objective and the constraints and to quantify the uncertainty.
We explain the details of this iterative procedure in the remainder of this section. 

Assuming that we are given an initial set of design parameters 
\begin{equation}
\Theta_{\init}:= \{\theta_i\in\Theta\ \! |\ \! i=1,\ldots,m_\init\}.    
\end{equation}
This set can be obtained either by random feasible perturbations of nominal design parameters, $\theta_{\text{nom}}$, or, one may employ Latin hyper-cube experiment design \cite{mckay2000comparison} to have maximally informative $\Theta_{\init}$.
The proposed tuning approach maintains a set of data $\Dcal_m$ defined as 
\begin{equation}
\Dcal_m:=\{(\theta_i,\xi_i^\sss{0},\xi_i^\sss{1},\xi_i^\sss{2})\ \!|\ \! i=1,\ldots,m\},    
\end{equation}
where $m$ is the iteration index, $\theta_i$ is the design parameters corresponding to the $i^\nth$ experiment, and, for $j=0,1,2$, $\xi_i^\sss{j}$ denotes the computed value for $g_j(\theta_i)$ based on the measured data in the $i^\nth$  experiment.
More precisely, we define $\xi_i^\sss{j}=g_j(\theta_i)+n_i^\sss{j}$, for all $i$ and $j$,
where $n_i^\sss{j}$ is a noise term introduced to capture the impact of uncertainty in the data collection procedure.
At iteration $m\ge m_\init$,  
using data $\Dcal_m$ and Gaussian process regression (GPR), for $j=0,1,2$, we build a surrogate function for $g_j$, denoted by $\hat{g}_{m}^\sss{j}$.
More precisely, let $\GP(\mu^\sss{j},k^\sss{j})$ be a Gaussian process prior for function $g_j$, where $\mu^\sss{j}:\Theta\to\Rbb$ and $k^\sss{j}:\Theta\times\Theta\to\Rbb$ are respectively the mean and kernel functions in the GP prior taken for $g^\sss{j}$.
Define information matrices $\btheta_m:=[\theta_1,\ldots,\theta_m]^\transp$ and $\bxi_m^\sss{j}:=[\xi_1^\sss{j},\ldots,\xi_m^\sss{j}]^\transp$.
Subsequently, for any $\theta\in\Theta$, 
we have
\begin{equation}
g^\sss{j}(\theta)\sim\Ncal\left(\mu_m^\sss{j}(\theta),\nu_m^\sss{j}(\theta)\right), \qquad j=0,1,2,     
\end{equation}
with mean and variance defined as
\begin{align}
\label{eqn:post_fm}
\!\!\!\!
\mu_m^\sss{j}(\theta) &\!:=\! 
\mu^\sss{j}(\theta) 
\!+\! 
\vc{k}_{m}^\sss{j}(\theta)^\transp  
(\mx{K}_{m}^\sss{j} \!+\! \sigma_{j}^{2}\eye)^{-1} 
(\bxi_m^\sss{j}\!-\!\mu^\sss{j}(\btheta_m)),  \!\!
\\ \label{eqn:cov}
\!\!\!\!
\nu_m^\sss{j}(\theta) &\!:=\!
k^\sss{j}(\theta,\theta)
\!-\!
\vc{k}_{m}^\sss{j}(\theta)^\transp  
(\mx{K}_{m}^\sss{j} \!+\! \sigma_{j}^{2}\eye)^{-1} 
\vc{k}_{m}^\sss{j}(\theta), \!\!
\end{align}
where
$\eye$ denotes the identity matrix, $\sigma_{j}^2$ is the variance of uncertainty in the evaluations of $g_j(\theta)$, 
vector $\vc{k}_{m}^\sss{j}(\theta)$ and $\mu^\sss{j}(\btheta_m)$ are respectively defined as 
\begin{equation}
\vc{k}_{m}^\sss{j}(\theta):= 
[k^\sss{j}(\theta,\theta_1),\ldots,k^\sss{j}(\theta,\theta_m)]^\transp \in\Rbb^m,    
\end{equation}
and 
\begin{equation}
\mu^\sss{j}(\btheta_m) =
[\mu^\sss{j}(\theta_1),\ldots,\mu^\sss{j}(\theta_m)]^\transp \in\Rbb^m,   \end{equation}
and matrix $\mx{K}_{m}^\sss{j}$ is defined as $[k^\sss{j}(\theta_{i_1},\theta_{i_2})]_{i_1,i_2=1}^m \in\Rbb^{m\times m}$.

Using these probabilistic surrogate models for functions $g_j$, $j=0,1,2$, we can select the parameters $\theta_{m+1}\in\Theta$, to be used in the next experiment and in subsequent evaluations of $\xi_i^\sss{0}$, $\xi_i^\sss{1}$ and $\xi_i^\sss{2}$.  
As an \emph{acquisition function} encoding an exploration-exploitation trade-off, aiming for $\theta$ with highest expected improvement of the cost, we define the \emph{expected improvement} function, $a_{\text{EI},m}:\Theta\to \Rbb$, as
\begin{equation}\label{eq:EI}
a_{\text{EI},m}(\theta)
:=
\big{(}\eta_m(\theta)\Phi(\eta_m(\theta))
+
\varphi(\eta_m(\theta))\big{)}\ \! \nu_m^\sss{0}(\theta)^{\frac{1}{2}} \, ,
\end{equation}
where
$\Phi$ and $\varphi$ respectively denote the cumulative distribution function 
and the probability density function 
of the standard normal distribution, and $\eta_m(\theta)$ is defined as
\begin{equation}
\eta_m(\theta):=(\mu_m^\sss{0}(\theta)-\xi_m^{0,+})/\nu_m^\sss{0}(\theta)^{\frac{1}{2}},    
\end{equation}
where $\xi_m^{0,+}$ denotes the minimal cost observed amongst the first $m$ experiments.
For considering the constraints,  the \emph{constrained expected improvement}  \cite{pmlr-v32-gardner14} is defined as
\begin{equation}\label{eq:CEI}
a_\text{CEI,m}(\theta)=  a_\text{EI,m}(\theta)\ \!
\prod_{j=1,2}\!\!
\Phi\left(
\frac{q_{j,\text{tr}}-\mu_m^\sss{j}(\theta)}{\nu_m^\sss{j}(\theta)^{\frac{1}{2}}}\right).
\end{equation}
Note that the second term in $a_\text{CEI,m}(\theta)$ indicates the feasibility probability of the design parameters $\theta$. 
To design the next vector of parameters, we solve the following optimization problem
\label{eqn:x_next} using global optimization heuristics such as \emph{particle swarm} optimization
\begin{equation}
\theta_{m+1} := \argmax_{\theta\in\Theta}\ \! a_\text{CEI,m}(\theta) \, .
\end{equation}
Once $\theta_{m+1}$ is obtained, we perform a new experiment in which the design parameters are set to $\theta_{m+1}$, and subsequently, the values of $\{g_j(\theta_{m+1})\}_{j=0}^2$ are evaluated and added to the data set
\begin{equation}
\Dcal_{m+1} = \Dcal_m \cup\{(\theta_{m+1},\xi_{m+1}^\sss{0},\xi_{m+1}^\sss{1},\xi_{m+1}^\sss{2})\} \, ,
\end{equation}
and the process is repeated.
This iterative procedure generates a sequence of $\{\theta_m|m=1,2,\ldots\}$ which converges to the solution of \eqref{eq:BO_min_f_st_g1g2} \cite{pmlr-v32-gardner14}.

\begin{figure}[t]
    \centering
    \includegraphics[width=0.45\textwidth]{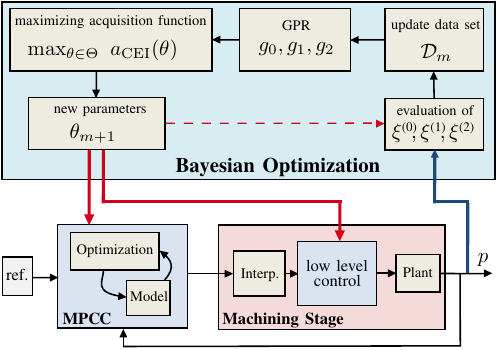}
    \caption{The scheme of data-driven tuning of MPCC based on Bayesian optimization}
    \label{fig:BO_MPCC}
\end{figure}

Figure \ref{fig:BO_MPCC} summarises the  introduced  Bayesian optimization-based scheme for data-driven tuning of MPCC.

We use squared-exponential kernels for $k^\sss{j}$, $j=0,1,2$, due to their universality property. 
Using  $\Theta_{\text{init}}$, the kernels' hyperparameters are tuned by minimizing the negative log marginal likelihood \cite{Rasmussen}, and remain unchanged during the introduced tuning procedure.
While re-tuning the hyperparameters of kernels at each iteration might improve the Gaussian process models, this may lead to disturbing the convergence of the introduced scheme. In the practical implementation, we set a stopping condition for the introduced optimization algorithm. To this end, one can consider various criteria such as a bound for maximal number of iterations (as in our implementation), or observing consecutive repeated solutions of \eqref{eqn:x_next} with approximately minimal observed cost \cite{khosravi2020performance}.
We set the mean functions as  $\mu^\sss{j}=0$, $j=0,1,2$ \cite{Rasmussen}. However, if we are given some prior information on the shape and structure of $g_j$, $j=0,1,2$, e.g., from similar experiments or numerical simulations, we can employ other options for the mean functions \cite{Rasmussen}.

\section{Results}
\label{sec:res}

We use two geometries to evaluate the performance of the proposed approach, an octagon geometry with edges in multiple orientations with respect to the two axes, and a curved geometry (infinity shape) with different curvatures, shown in Figure \ref{fig:noMPCinf}. We have implemented the simulations in Matlab, using Yalmip/Gurobi to solve the corresponding MPCC quadratic program in a receding horizon fashion and the GPML library for Gaussian process modeling. We compare three schemes: manual tuning of the MPCC parameters for fixed low level controller gains, Tuning of MPCC parameters  through Bayesian optimization, and joint tuning of the MPCC- and the low-level cascade controller parameters using Bayesian optimization.
\vspace{-2mm}
\subsection{Manual tuning of the MPCC parameters}

We first optimize the performance of the simulated positioning system by adding a receding horizon MPCC stage where we pre-optimize the position and velocity references provided to the low level controller. This is enabled by the high repeatability of the system which results in run-to-run deviations of $3 \mu m$, well below our tolerances of $20 \mu m$. The weights in the MPCC cost terms are manually tuned, the controller gains are kept at their nominal values, and the horizon length is set to $25$ time steps.

\begin{figure}[h]
    \centering
    \includegraphics[width=0.45\textwidth]{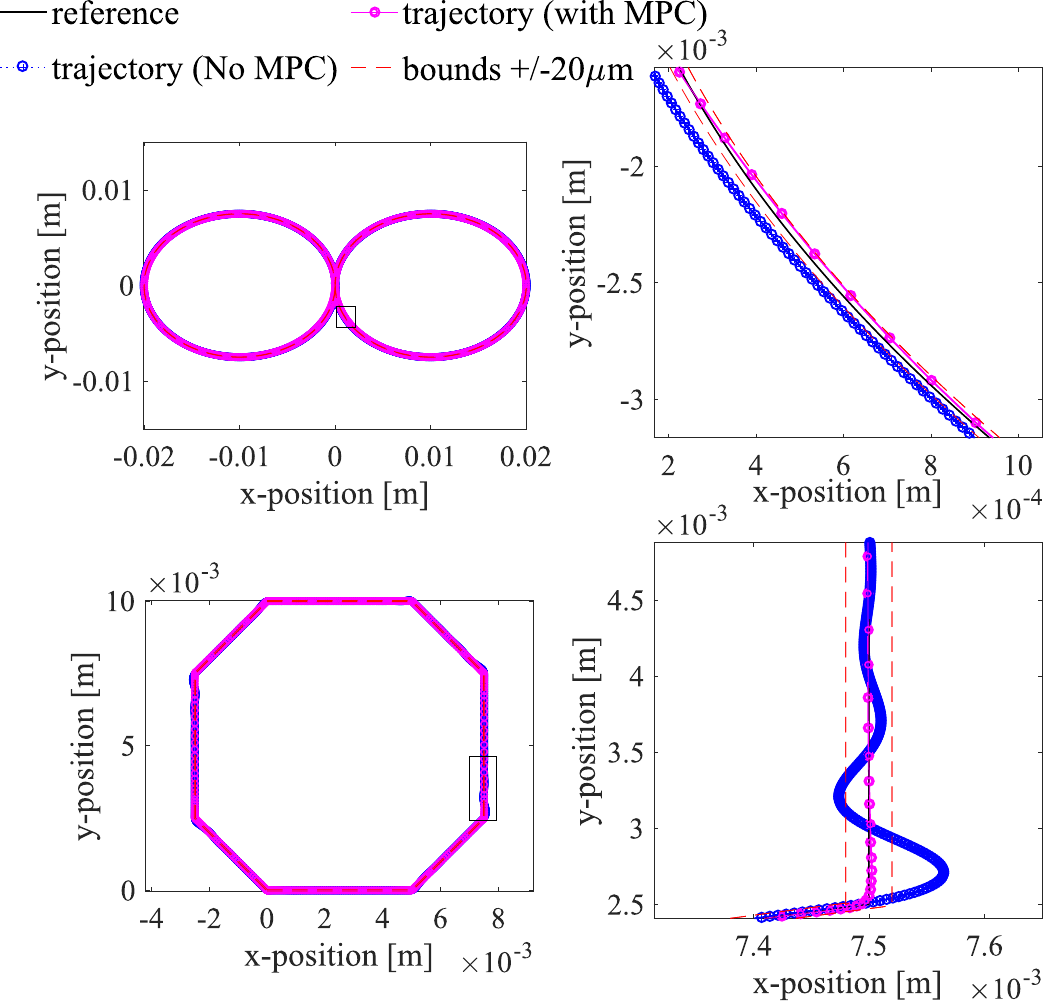}
    \caption{ Experimental results for the two geometries showing the tracking performance with the nominal controller, and with the MPC-based planner with manually tuned parameters.}
    \label{fig:noMPCinf}
\end{figure}

Table \ref{tab:performance_nominal} shows the nominal performance giving the reference geometry directly as an input to the existing controller compared to the performance after adding the manually-tuned MPCC. The two error metrics are defined as follows: $\|\hat{e}_c\|_\infty:=\|[\hat{e}_{c,k}]_{k=1}^{k_{\mathrm{tot}}}\|_{\infty}$, and $\|\hat{e}_c\|_2:=\|[\hat{e}_{c,k}]_{k=1}^{k_{\mathrm{tot}}}\|_2$, over the whole geometry. Adding the MPCC results in a a 9-fold improvement in the traversal time, along with maintaining the maximum deviation $\|\hat{e}_c\|_\infty$ within the imposed bounds of $ \pm 20 \mu m$. Both the average and the maximal tracking errors exhibit a 3-fold decrease. The significant improvement in tracking following pre-optimization with the MPCC is illustrated on Figure \ref{fig:noMPCinf} for the two geometries.

\begin{table}
\caption{Nominal performance over the whole geometry}
\label{tab:performance_nominal}
\centering 
\begin{tabular}[h]{@{}l c c c c c @{}}\toprule
& \multicolumn{2}{c}{infinity} &&  \multicolumn{2}{c}{octagone}\\
\cmidrule{2-3} \cmidrule{5-6}
 & nominal & MPC &&  nominal & MPC \\
 \midrule
$\|\hat{e}_c\|_2$ [$\mu$m]   & 13.5 & 3.8 & & 5.7 & 0.78 \\
$\|\hat{e}_c\|_\infty$ [$\mu$m] & 60.1 & 20 & & 65.3 & 19.6\\
Maneuver time [s]   & 9.21 & 1.29 & & 8.53 & 1.22 \\ 

\bottomrule
\end{tabular}
\end{table} 

\subsection{Bayesian Optimization of the MPCC parameters}

 \begin{figure*} [h]
\begin{subfigure}[c]{0.23\textwidth}
\includegraphics[width=1\textwidth]{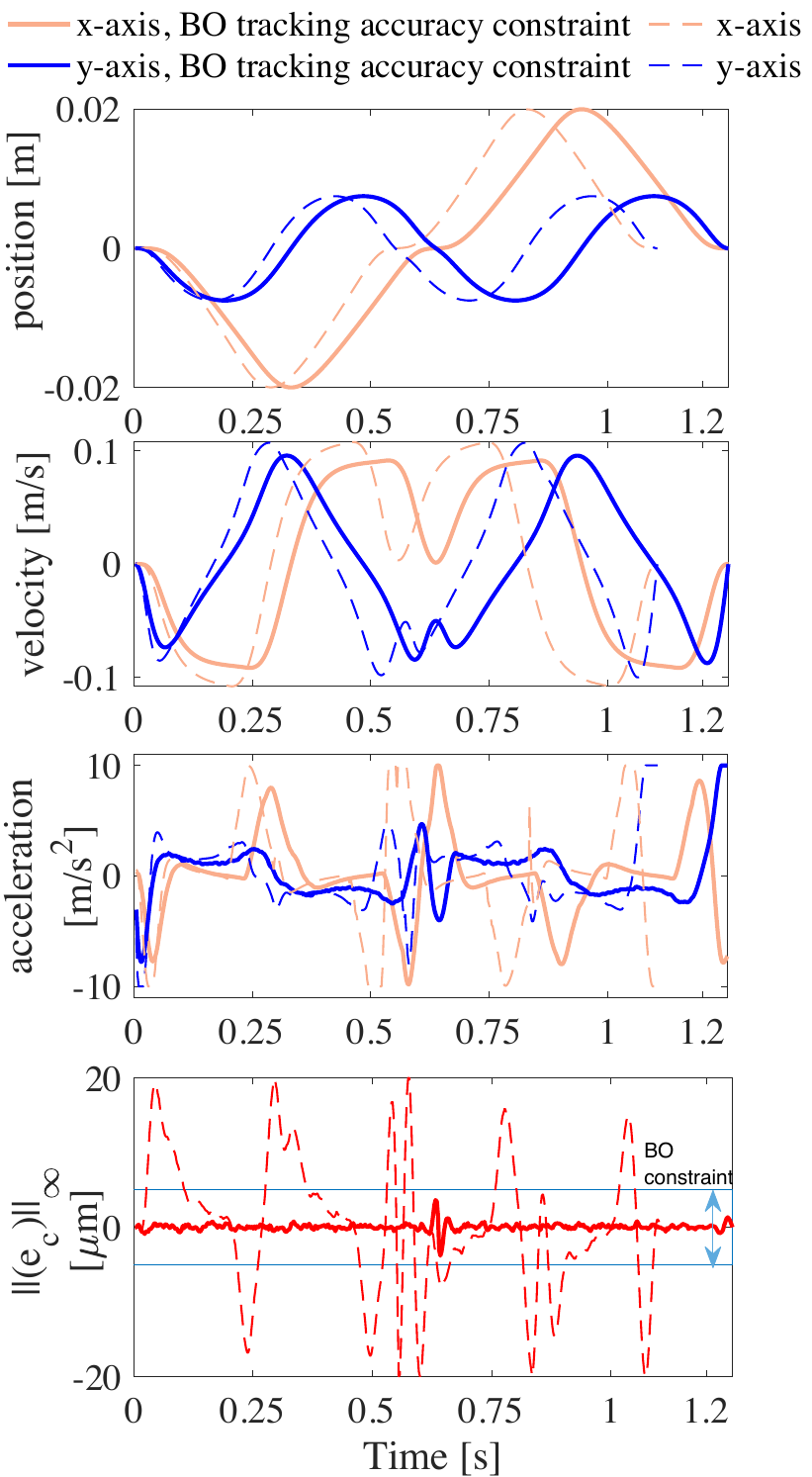}
\label{fig:MOC_BO_inf}
\end{subfigure}
\hfill
\hspace{-3mm}
\begin{subfigure}[c]{0.23\textwidth}
\includegraphics[width=1\textwidth]{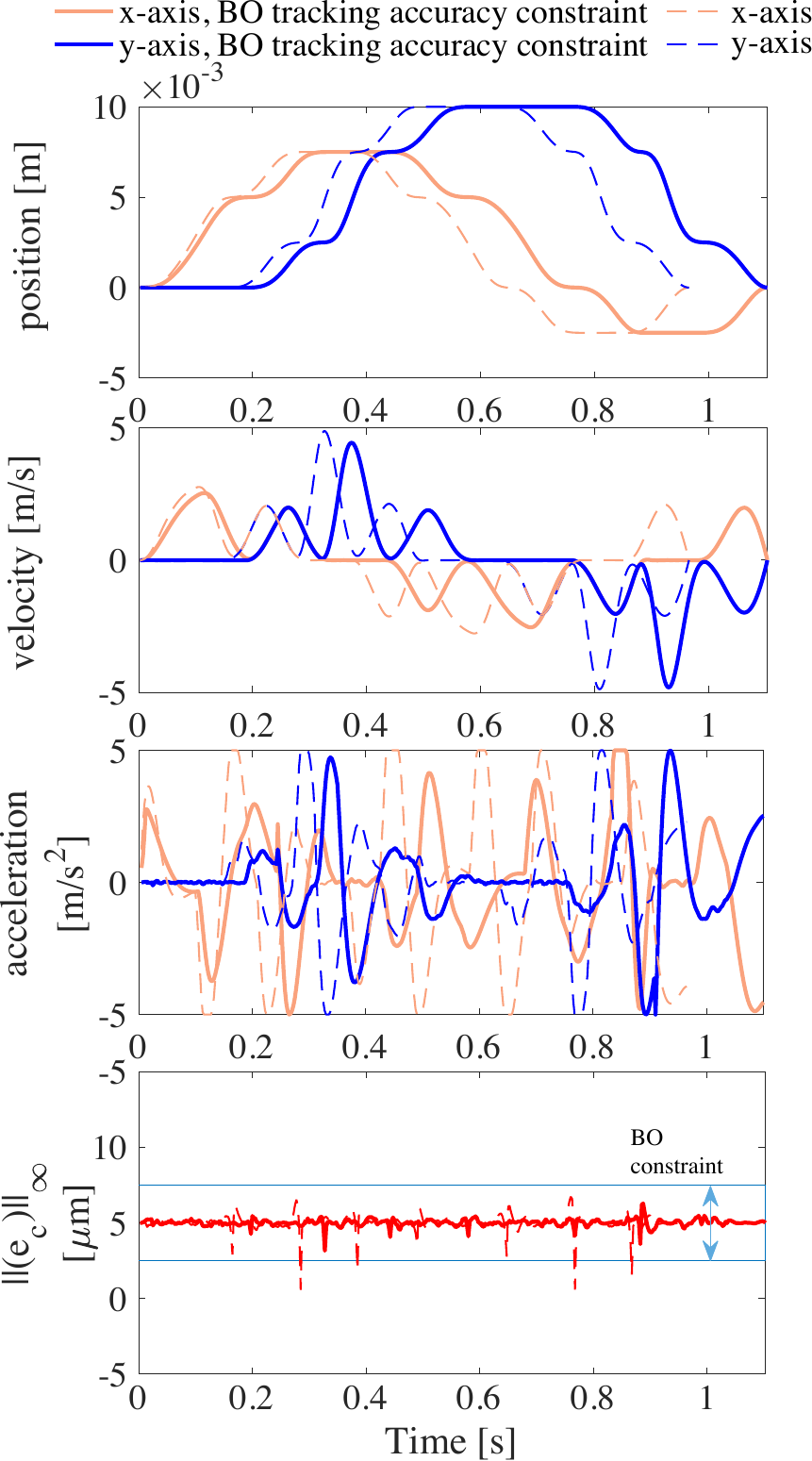}
\label{fig:MOC_BO_inf}
\end{subfigure}
\hfill
\hspace{-3mm}
\begin{subfigure}[c]{0.23\textwidth}
\includegraphics[width=1\textwidth]{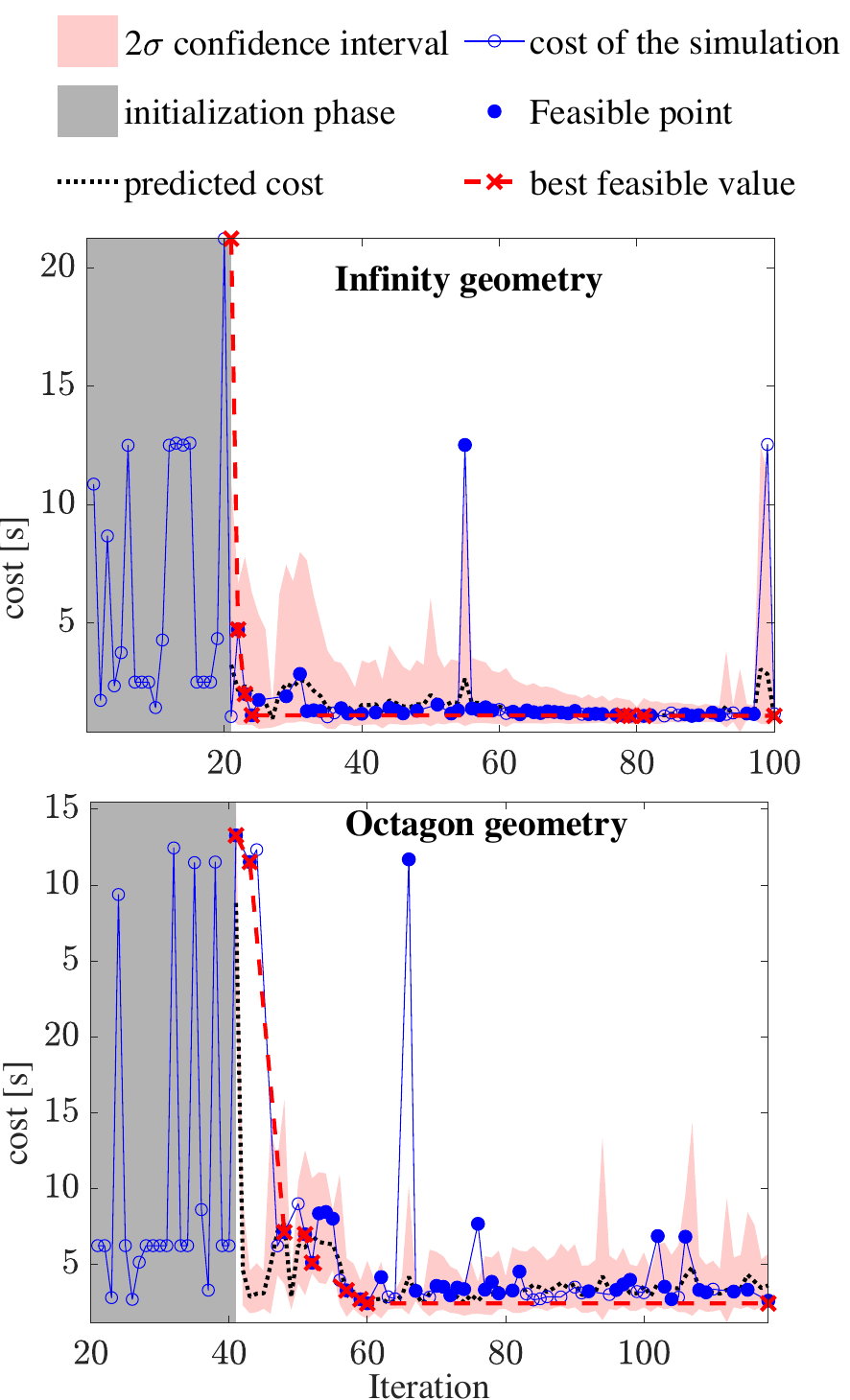}
\label{fig:MOC_BO_progress}
\end{subfigure}
\hfill
\hspace{-6mm}
\footnotesize\addtolength{\tabcolsep}{-6pt}
    \raisebox{0.3cm}{
    \begin{tabular}{ @{}l c c c c c @{} }
    \toprule
   & time  & $\|\hat{e}_c\|_\infty$ & $\|\hat{e}_c\|_2$ \\
   & $[s]$ & $[\mu m]$ & $[\mu m]$ 
   \\\midrule\midrule
   BO - MPCC  
   \\
   (fixed low-level par.)  
   \\\midrule
    infinity & 1.11 & 20 & 7.4\\
    oct.  & 0.97 & 8.9 & 0.55\\\midrule
    BO - MPCC $\|\hat{e}_c\|_\infty$ constr. &  &  &  \\
    (fixed low-level par.)   &  &  &  \\\midrule
       infinity &1.259 & 3.78 & 0.29\\
    oct.  & 1.12 & 7.8  & 0.29\\ \midrule
     BO - low level par. &  &  &  \\
     (fixed MPCC par.) &  \\\midrule
    infinity & 0.93 & 20 & 3.9\\
    oct.  & 0.97 & 20 & 1.17\\\midrule
   \bottomrule
\end{tabular}}
\vspace{-2mm}
\caption{Position, velocity, acceleration, and maximal contour error resulting from optimization of the MPC parameters, comparing unconstrained BO optimization (solid lines) to BO optimization with additional constraint on the maximal tracking error, for infinity (left) and octagon(center) geometries. The right panel shows the evolution of BO iterations, until optimization terminates. The performance metrics evaluating infinitytracking accuracy and time are summarized in the table for unconstrained and constrained BO.}
\vspace{-0.3cm}
\label{fig:BO_MPC_summary}
\end{figure*}
 
To automate the tuning of the MPCC, we use the approach outlined in \eqref{eq:BO_min_f_st_g1g2}. We optimize the MPCC parameters as specified in \eqref{eq:opt_var}, while keeping the low level controller gains fixed. We introduce a safety constraint following \eqref{eq:BO_min_f_st_g1g2},  with $q_{1,\mathrm{tr}}=2000 m/s^3$. To improve further the tracking, we introduce an additional tracking error constraint with $q_{2,\mathrm{tr}}= 5 \mu m$. The resulting trajectories for the position, velocity, and acceleration input of each axis are shown in Figure \ref{fig:BO_MPC_summary}.

As expected, adding the global tracking error constraint increases the traversal time, but maintains the maximal deviation within the bounds (see the table in \ref{fig:BO_MPC_summary}). This tracking error constraint results in a dramatic 5-fold decrease of the maximum deviation $\|\hat{e}_c\|_\infty$, at the cost of an increase of the traversal time only by 10\%, and the traversal time is still better compared to the one achieved with manually tuned MPCC. The constraints on the jerk successfully result in avoiding big jumps in the acceleration. 

For the initialization phase needed to train the GPs in the Bayesian optimization, we select 20 samples over the whole range of MPC parameters, using Latin hypercube design of experiments. The BO progress is shown in Figure \ref{fig:BO_MPC_summary}, right pannel, for the optimization with constraints on the jerk and on the tracking error. After the initial learning phase the algorithm quickly finds the region where the simulation is feasible with respect to the constraints. The confidence interval in the cost prediction narrows for the infinity shaped trajectory, which is likely due to a more clear minimum in the cost of this geometry. The optimization stops after a fixed number of iterations is reached, and the parameters are set to those corresponding to the best observed cost.

\section{Conclusion}
\label{sec:conc}
This paper demonstrated a hierarchical contour control implementation for the increase of productivity in positioning systems. We use a contouring predictive control approach to optimize the input to a low level controller. This control framework requires tuning of multiple parameters associated with an extensive number of iterations. We propose a sample-efficient joint tuning algorithm, where the performance metrics associated with the full geometry traversal are modelled as Gaussian processes, and used to form the cost and the constraints in a constrained Bayesian optimization algorithm, where they enable the trade-off between fast traversal, high tracking accuracy, and suppression  of vibrations in the system. Data-driven tuning of all the parameters compensates for model imperfections and results in improved performance.
Our numerical results demonstrate that  tuning the  parameters of the MPCC stage achieves the best performance in terms of time and tracking accuracy.

\section{Acknowledgements}

We thank Timo Roth for his support in parts of the numerical implementation.

\bibliography{references}
\bibliographystyle{IEEEtran}
\end{document}